\documentclass[aps,pre,twocolumn]{revtex4}
\usepackage[pdftex]{graphicx}

\begin{document}
\title{Liquid-theory analogy of direct-coupling analysis of multiple-sequence alignment and its implications for protein structure prediction}
\author{Akira R. Kinjo\\
  Institute for Protein Research, Osaka University, \\
  3-2 Yamadaoka, Suita, Osaka, 565-0871, Japan}

\begin{abstract}
  The direct-coupling analysis is a powerful method for protein contact prediction, and enables us to extract ``direct'' correlations between distant sites that are latent in ``indirect'' correlations observed in a protein multiple-sequence alignment. I show that the direct correlation can be obtained by using a formulation analogous to the Ornstein-Zernike integral equation in liquid theory. This formulation intuitively illustrates how the indirect or apparent correlation arises from an infinite series of direct correlations, and provides interesting insights into protein structure prediction. 
\end{abstract}
\maketitle

\section{Introduction}
Protein multiple-sequence alignments (MSA) are a useful means to extract various and valuable information about protein families~\cite{DurbinETAL}. It is well recognized that the frequency of amino acid residues at each alignment site is a useful measure of its functional importance. It has also been suggested that correlation between distant sites along the sequence is a rich source of information about the structure and function of the protein families\cite{Toh2004}.
In fact, recent years have seen a significant advance in our understanding of the site-site correlation observed in MSA. Of particular importance is the development of direct-coupling analysis (DCA) and related methods\cite{MorcosETAL2011,JonesETAL2012,Miyazawa2013}. Although the basic idea has been already suggested in the last century\cite{LapedesETAL1999}, it is only by the recent explosion of protein sequence data, in addition to theoretical development, that practical implementation of the idea was made possible. What DCA tells us is clear: The ``apparent'' correlation  observed in a MSA is a result of ``direct'' correlations which are closely related to structural contacts. For example, if residues $i$ and $j$ are in physical contact (directly correlated), and so are residues $j$ and $k$, then residues $i$ and $k$ may appear to be correlated even if they are not in contact.

There are many variants of DCA today. A major one is based on the principle of maximum entropy\cite{MorcosETAL2011}, others are based on the graphical Gaussian model\cite{JonesETAL2012} or phylogenetic analysis\cite{Miyazawa2013}.
All of these methods are good predictors of physical contacts between residues in native protein structures.
In this Note, I derive the direct correlation based on a formulation that is analogous to the integral equation theory of simple liquids\cite{SimpleLiquids3}. This formulation has an advantage in that it intuitively shows how apparent correlations are realized by an infinite series of direct correlations. Based on the analogy with the liquid theory, it may be possible to elaborate the theory of direct correlations in MSA. More importantly, the intuitive picture that the present analysis provides helps us examine the mechanism of protein structure prediction from a new perspective, which may in turn lead to the development of new methods based on novel principles.

\section{Theory}
A multiple-sequence alignment consisting of $M (\gg 1)$ amino acid sequences and $N$ alignment sites may be regarded as an $M\times N$ matrix of symbols. That is, each row represents an amino acid sequence including gap symbols and each column represents an alignment sites. Let $n_{k,i}(a) = 1$ if the residue type $a$ appears at the site $i$ of the sequence $k$, otherwise let $n_{k,i}(a) = 0$. We first define the frequency $n_{i}(a)$ of residue $a$ at site $i$ as
\begin{equation}
  n_{i}(a) = \frac{1}{M}\sum_{k=1}^{M}n_{k,i}(a).
\end{equation}
Next, the correlation (covariance) between residue $a$ at site $i$ and residue  $b$ at site $j$ is defined as
\begin{equation}
  C_{ij}(a,b) = \frac{1}{M}\sum_{k=1}^{M}[n_{k,i}(a) - n_{i}(a)][n_{k,j}(b) - n_j(b)].
\end{equation}
For simplicity, we assume that there are a sufficient number of sequences so that these statistics can be computed sufficiently accurately, and also ignore the effect of the phylogenetic bias in a family of sequences. Another caveat is required when there are completely conserved sites in the case of which the columns and rows corresponding to those conserved sites are zero. We assume this problem is properly taken care of, for example, by adding pseudo-counts.
The correlations as a whole can be regarded as a $21N\times 21N$ matrix by properly ordering residues and sites. Note that, since the equality $\sum_{a=1}^{21}n_{k,i}(a) = 1$ holds for any sequence $k$, the matrix $C$ is rank-deficient. Nevertheless, it can be made invertible by removing the rows and columns corresponding to the gap symbol, and hence the size of the matrix $C$ is now $20N\times 20N$, which is assumed in the following.

\begin{figure}
  \begin{center}
  \includegraphics[width=8cm]{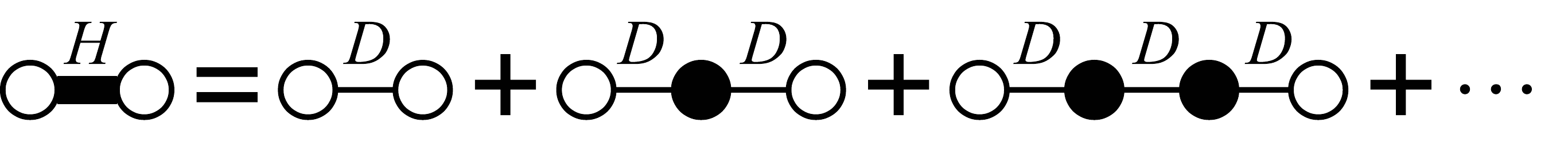}    
  \end{center}
  \caption{\label{fig:dia}A diagrammatic representation of Eq. \ref{eq:oz} with $H = C - \rho$.}
\end{figure}
Now we assume there exists a ``direct correlation'' $D_{ij}(a,b)$ between residue $a$ at site $i$ and residue $b$ at site $j$, and the correlation $C$ is a result of an infinite series of the direct correlations:
\begin{eqnarray}
  C_{ij}(a,b) &=& n_{i}(a)\delta_{i,j}\delta_{a,b} 
   + n_{i}(a)D_{ij}(a,b)n_{j}(b) \nonumber\\
  &&+  \sum_{k,c}n_{i}(a)D_{ik}(a,c)n_{k}(c)D_{kj}(c,b)n_{j}(b) \nonumber\\
  &&+  \sum_{k_1,c_1}\sum_{k_2,c_2}n_{i}(a)D_{ik_1}(a,c_1)n_{k_1}(c_1)\nonumber\\
  &&~~\times D_{k_1k_2}(c_1,c_2)n_{k_2}(c_2)D_{k_2j}(c_2,b)n_{j}(b) \nonumber\\
  &&+ \cdots
\end{eqnarray}
By defining the diagonal matrix $\rho_{ij}(a,b) = n_{i}(a)\delta_{i,j}\delta_{a,b}$, this equation is expressed as
\begin{eqnarray}
  C &=& \rho + \rho D \rho + \rho D \rho D \rho + \rho D \rho D \rho D \rho + \cdots\label{eq:oz}\\
  &=& \rho + \rho D C.
\end{eqnarray}
This matrix equation is analogous to the Ornstein-Zernike integral equation in the theory of simple liquids\cite{SimpleLiquids3} and can be expressed as a diagram in Figure \ref{fig:dia} (where the left-hand side represents $H = C - \rho$). By solving this equation for $D$, we have
\begin{equation}
  D = \rho^{-1} - C^{-1}\label{eq:dc}
\end{equation}
which is essentially equivalent to the result of the mean-field DCA derived by Morcos et al.\cite{MorcosETAL2011} based on the Plefka expansion\cite{Plefka1982}.

\section{Discussion}
While Morcos et al.\cite{MorcosETAL2011} used direct correlations as \emph{pairwise interactions} between residues, direct correlations (in liquid theory) are generally different from interactions. In fact, the approach of Morcos et al. may be interpreted as the mean-spherical approximation\cite{SimpleLiquids3} which is a particular closure condition for solving the Ornstein-Zernike equation. It may be interesting to investigate  other choices of closure conditions such as those analogous to, for example, the Percus-Yevick (PY) or hypernetted-chain (HNC) approximations\cite{SimpleLiquids3}. The HMSA closure\cite{ZerahANDHansen1986} is another interesting possibility.

By rearranging Eq. (\ref{eq:dc}), we have
\begin{equation}
  \rho = \left(D + C^{-1}\right)^{-1}. \label{eq:rhodef}
\end{equation}
This relation can be interpreted as a self-consistent condition (rather than a ``definition'') for $\rho$ when $D$ is given, and shows how the position-specificity of residue frequencies depends on the entire context of a protein family and its structure.
It is now widely accepted that sequence-based profile methods\cite{AltschulETAL1997,KroghETAL1994} are the best method for template-based structure prediction. Noting that the direct correlations well correspond to native contacts, Eq. (\ref{eq:rhodef}) tells us that an infinite series of tertiary interactions are effectively convoluted into a sequence profile through the alignment of many evolutionarily related sequences.
On the contrary, purely structure-based profile or threading methods\cite{ProteinBioinfo}, intuitively speaking, take into account only the first one or two terms in Eq. (\ref{eq:oz}) where $\rho$ in this case is position-\emph{independent}. This may be a reason for the insufficient position-specificity, and hence the limited success, of purely structure-based profile methods.

The present analysis also has an implication for template-free or \emph{de novo} structure prediction. All template-free methods are based on some empirical energy or scoring functions (whether physicochemical or statistical) and suffer from the problem of a rugged energy landscape that leads to many suboptimal non-native structures. In the mean time, studies on protein folding have shown that the energy landscape of natural proteins is minimally frustrated and funnel-like.
This property can be readily modeled by the Go-like potentials in which only the native contacts are stabilizing\cite{Go1983,OnuchicANDWolynes2004}. It is conjectured that natural proteins have been naturally selected to satisfy such property  in the course of molecular evolution\cite{Go1983}. This observation suggests a way to improve structure prediction by improving protein sequence design. That is, an empirical energy function that can reproduce the sequence profiles of (natural) protein families in the (re)designing process (i.e., generating sequences compatible with a given native structure)\cite{KuhlmanANDBaker2000,OllikainenETAL2013} may be expected to realize the ``correct'' direct correlation and development of such an energy function may help improve structure prediction.

Physicochemically, it is the sequence that determines the structure. Evolutionarily, however, it is the structure that molds the pattern of a family of sequences. The DCA sheds new light especially on the latter aspect of proteins by explicitly providing the relation between the observed correlation $C$ (i.e., the pattern of sequences) and the direct correlation $D$ ($\approx$ physical contacts). I hope the present analysis help further clarify the meaning of this intricate relationship between protein sequences and structures.

\begin{acknowledgments}
I thank Mr. Iseo Nose whose lecture on Goethe's morphology motivated me to write this note.
\end{acknowledgments}


\bibliography{refs,mypaper}


\end{document}